\documentclass[10pt,aps,showpacs,pra,twocolumn,superscriptaddress]{revtex4-2}

\usepackage[colorinlistoftodos, textsize=small]{todonotes}

\usepackage{graphicx}
\usepackage{times}
\usepackage{amsfonts,amsthm}
\usepackage{physics}
\usepackage{amssymb}
\usepackage{mathtools}
\usepackage[utf8]{inputenc}
\usepackage[polish,english]{babel}
\usepackage{MnSymbol}
\usepackage[T1]{fontenc}
\usepackage{xcolor}
\let\ChangesComment\comment
\usepackage{changes}
\let\comment\ChangesComment
\usepackage{comment}
\usepackage[colorlinks=true,citecolor=blue,urlcolor=blue]{hyperref}

\newtheorem{thm}{Theorem}
\newtheorem{defn}[thm]{Definition}

\begin{document}

\title{Equivalence of genuine multipartite entanglement and nonlocality of nearly symmetric multiqubit pure states}

\author{Jakub Wójcik}
\email{j.wojcik32@student.uw.edu.pl}
\affiliation{Center for Theoretical Physics, Polish Academy of Sciences, Aleja Lotnik\'{o}w 32/46, 02-668 Warsaw, Poland}
\author{Wojciech Bruzda}
\email{w.bruzda@cft.edu.pl}
\affiliation{Center for Theoretical Physics, Polish Academy of Sciences, Aleja Lotnik\'{o}w 32/46, 02-668 Warsaw, Poland}
\author{Ignacy Stachura}
\affiliation{Center for Theoretical Physics, Polish Academy of Sciences, Aleja Lotnik\'{o}w 32/46, 02-668 Warsaw, Poland}
\author{Remigiusz Augusiak}
\affiliation{Center for Theoretical Physics, Polish Academy of Sciences, Aleja Lotnik\'{o}w 32/46, 02-668 Warsaw, Poland}

\date{\today}

\begin{abstract}
Whether every pure genuinely multipartite entangled (GME) state necessarily exhibits genuine multipartite nonlocality (GMNL) remains an open question. By combining a recently proposed Bell inequality [I. Stachura \textit{et al.}, \href{https://iopscience.iop.org/article/10.1088/1367-2630/ad7753}{New J. Phys. \textbf{26}, 093029 (2024)}] with Hardy's paradox and the canonical decomposition of pure states, we analytically demonstrate that all highly symmetric, genuinely entangled multipartite qubit states exhibit genuine multipartite nonlocality, thereby supporting Gisin's conjecture in the multipartite setting. This result constitutes a step toward a general proof of the conjectured equivalence between GME and GMNL in quantum theory.
\end{abstract}

\maketitle

\section{Introduction}

Quantum entanglement is a distinctive feature of quantum mechanics, representing a class of correlations that cannot be accounted for by any classical theory. In a bipartite system, a pure state $|\psi\rangle\in\mathcal{H}_A\otimes\mathcal{H}_B$ is called entangled if it cannot be written as a product of states of its subsystems, i.e., if $|\psi\rangle\neq|\psi_A\rangle\otimes|\psi_B\rangle$. Mixed-state entanglement is similarly defined by the inability to express the state as a probabilistic mixture of such product states~\cite{PhysRevA.40.4277} (cf. also \cite{HHHH09}). In multipartite systems, where more than two parties are involved, entanglement exhibits a more complex structure~\cite{HRZ24}. In this setting, it is natural to ask whether the entanglement involves all parties simultaneously, or only particular smaller groups. This leads to the notion of genuine multipartite entanglement (GME) which describes the lack of separability with respect to any bipartition of the system (formal definitions will follow in the next section) \cite{PhysRevLett.83.3562,Dur_2000,Dur2_2000,Acin_2001}. Genuine multipartite entanglement is thus the strongest form of entanglement in a multipartite system as it cannot be attributed to any subset of parties, involving all of them simultaneously. At the same time, GME represents a valuable resource in the multipartite regime with applications ranging from quantum metrology \cite{Tóth_2014} and multipartite secret sharing \cite{PhysRevA.59.1829,PhysRevA.78.042309} to measurement-based quantum computing \cite{PhysRevLett.86.5188}. A paradigmatic example of a GME state is the three-qubit Greenberger-Horne-Zeillinger state~\cite{GHSZ90}:
\begin{equation}
|\mathrm{GHZ}_3\rangle=\frac{1}{\sqrt{2}}\left(|000\rangle+|111\rangle\right).
\end{equation}
This state cannot be written as a product across any bipartition of the three qubits, and it exhibits correlations that involve all three parties in an irreducible way. It is therefore genuinely tripartite entangled.

While any state exhibiting Bell nonlocality must be entangled, not all entangled states violate Bell inequalities (BI) as there exists examples of mixed states both in the bipartite \cite{PhysRevA.40.4277,PhysRevA.65.042302}  and multipartite~\cite{PhysRevLett.115.030404,Hirsch} cases for which local hidden variable models can be constructed. The situation is somewhat simpler in the case of pure states. In fact, the key result from Ref.~\cite{GISIN1991201,GISIN199215} states that any pure entangled bipartite state necessarily violates a Bell inequality -- equivalently, this implies the genuine multipartite nonlocality (GMNL) of any bipartite GME states. To some extent this observation was later generalized to the multipartite scenario in Ref.~\cite{POPESCU1992293}, where it was shown that any multipartite entangled state exhibits some form of nonlocality. Yet, a truly meaningful question in the multipartite regime is whether all genuinely entangled pure states are also genuinely nonlocal. The validity of this conjecture appears plausible, as it is supported by several examples:
in Refs.~\cite{Yu2013} (see also Ref.~\cite{Stachura_2024}) it has been proven that all three-qubit GME states are GMNL, whereas in Ref.~\cite{PhysRevLett.112.140404} all symmetric GME states of arbitrary number of qubits have been shown to exhibit GMNL. Furthermore, a recent study~\cite{Makuta2023} demonstrates that all genuinely entangled stabilizer subspaces are fully nonlocal and thus also genuinely nonlocal in the multipartite scenario. In other words, all pure multipartite states from GME stabilizer subspaces or all mixed states acting on them are GMNL. Despite this progress, the fundamental question of whether for pure states GME necessarily implies GMNL in full generality remains open.

In our work we establish the equivalence between GME and GMNL for a class of multipartite nearly symmetric states consisting of an arbitrary number of qubits. More precisely, we demonstrate that any genuinely entangled state from the Hilbert space $\mathbb{C}^2\otimes \mathrm{Sym}((\mathbb{C}^2)^{\otimes (n-1)})$, where $\mathrm{Sym}((\mathbb{C}^2)^{\otimes n})$ denotes the symmetric subspaces of $n$ qubits, exhibits genuine nonlocality. We thus generalize the result of Ref.~\cite{curchod} for three-qubit GME states with permutationally invariant last two parties, and that of Ref.~\cite{PhysRevLett.112.140404}, to an arbitrary number of parties and a broader class of states, respectively. The approach we follow to prove the above statement exploits a certain class of Bell inequalities detecting GMNL in multipartite systems introduced recently in Ref.~\cite{Stachura_2024}, which improve some of the inequalities obtained in an earlier work~\cite{curchod}.

The remainder of this paper is organized as follows. In Section~\ref{sec:concepts}, we introduce the formal definitions of genuine multipartite entanglement and genuine multipartite nonlocality. We also review a specific Bell inequality and discuss how it can be employed to infer GMNL. Section~\ref{sec:main-result} presents the main result of this work, namely, the application of this Bell inequality to a particular class of $n$-qubit states. Finally, in Section~\ref{sec:conclusion}, we outline possible directions for extending the analysis to arbitrary $n$-qubit states.

\section{Concepts and Methods}\label{sec:concepts}

Let $[n]\equiv \{1,2,\ldots,n\}$. Consider $|\psi\rangle \in \mathcal{H} = \bigotimes_{j=1}^n \mathbb{C}^{d_j}$ be a pure state shared by $n\geqslant 2$ parties $\{A_j\}_{j=1}^n$. The state $|\psi\rangle$ is genuinely multipartite entangled (GME) if it is not separable with respect to any bipartition of the parties. Formally:

\begin{defn}
Given a nontrivial bipartition of the index set $[n]$ into two disjoint, non-empty subsets $S \subsetneq [n]$ and its complement $\bar{S} = [n] \setminus S$, define
\begin{equation}
\mathcal{H}_S \equiv \bigotimes_{j \in S} \mathbb{C}^{d_j} \qquad \text{and}\qquad \mathcal{H}_{\bar{S}} \equiv \bigotimes_{j \in \bar{S}} \mathbb{C}^{d_j}.
\end{equation}
Then, $|\psi\rangle$ is genuinely multipartite entangled if for any subset $S \subsetneq [n]$ such that $S \neq \emptyset$, the state $\ket{\psi}\neq \ket{\psi_S}\otimes\ket{\psi_{\bar{S}}}$ for any $\ket{\psi_S}\in\mathcal{H}_S$ and any $\ket{\psi_{\bar{S}}}\in\mathcal{H}_{\bar{S}}$. Equivalently,  $|\psi\rangle$ is GME, iff it is entangled across any bipartition $S\,|\,\bar{S}$.
\end{defn}

In the case of mixed states, the definition extends by considering convex combinations of bipartite-separable states across all bipartitions. As the present work is restricted to pure states, the full definition is omitted and can be found, e.g., in Ref.~\cite{Makuta2023}.

Let us now define genuine multipartite nonlocality. Suppose that each party $A_j$, holds a share of the state 
$\rho$. Each party can choose one of $m$ possible $d$-outcome measurements $M_x^{(j)}=\{ M^{(j)}_{a|x} \}_{a=0}^{d-1}$, where $x$ labels the measurement setting and $a=0,\ldots,d-1$ denotes the outcome. In this work we are concerned only with rank-one projective measurements and therefore we can represent the measurement operators $M_{a\,|\,x}^{(j)}$ as projections
\begin{equation}
M_{a|x}^{(j)}=|m_{a\,|\,x}^{(j)}\rangle\langle m_{a\,|\,x}^{(j)}|,
\end{equation}
on some pure states $|m_{a\,|\,x}^{(j)}\rangle$ from the corresponding Hilbert space. As a result of a Bell experiment, the parties observe correlations that are represented by a collection of probability distributions organized in a form of a vector
\begin{align}
    \mathbf{p}&\equiv \Big\{p(\mathbf{a}\,|\,\mathbf{x}) = p(a_1\ldots a_n\,|\,x_1\ldots x_n)\Big\},
\end{align}
where each individual $p(a_1\ldots a_n\,|\,x_1 \ldots x_n)$ is the probability of 
obtaining the outcome $a_j$ by party $A_j$ after they perform the $x_j^{\mathrm{th}}$ measurement and can be represented as
\begin{equation}
    p(a_1 \ldots a_n\,|\,x_1\ldots x_n)=\mathrm{Tr}
    \left(\bigotimes_{j}M_{a\,|\,x}^{(j)}\,\rho\right).
\end{equation}

In order to introduce the notion of GMNL in a generic $n$-party scenario let us consider again a nontrivial bipartition $S|\bar{S}$. We call correlations $\mathbf{p}$ bilocal with respect to the bipartition $S|\bar{S}$ if the probabilities admit the following decomposition
\begin{equation}
p(\mathbf{a}\,|\,\mathbf{x})=\sum_{\lambda}r_{\lambda}\,p_{S}(\mathbf{a}_{S}\,|\,\mathbf{x}_{S},\lambda)\,p_{\bar{S}}(\mathbf{a}_{\bar{S}}\,|\,\mathbf{x}_{\bar{S}},\lambda),
\end{equation}
where $\lambda$ are hidden variables (referred also to as shared randomness) distributed with probability distribution $r_{\lambda}$, and $p_{S}(\mathbf{a}_{S}\,|\,\mathbf{x}_{S},\lambda)$ and $p_{\bar{S}}(\mathbf{a}_{\bar{S}}\,|\,\mathbf{x}_{\bar{S}},\lambda)$ are probability distrubutions corresponding to the sets of parties $S$ and $\bar{S}$, respectively. We then call $\mathbf{p}$ bilocal if it can be written as a convex combination of correlations which are bilocal with respect to various bipartitions, i.e.,
\begin{equation}\label{TintodeVerano}
    p(\mathbf{a}\,|\,\mathbf{x}) = \sum_k p_k \sum_{\lambda} r_{\lambda}^{(k)}\, p_{S_k\,|\,\bar{S}_k}(\mathbf{a}\,|\,\mathbf{x}, \lambda),
\end{equation}
where $p_k \geqslant 0$, $\ \sum_k p_k = 1$, and 
$p_{S_k\,|\,\bar{S}_k}(\mathbf{a}\,|\,\mathbf{x}, \lambda)$ represent correlations which are local with respect to the bipartition $S_k|\bar{S}_k$.

\begin{defn}Correlations $\mathbf{p}=\{p(\mathbf{a}|\mathbf{x})\}$ are called truly multipartite entangled if there is no decomposition~\eqref{TintodeVerano} for them.
\end{defn}

\iffalse
\begin{defn}
\replaced{A given correlation $\mathbf{p}$}{A vector of correlations $\mathbf{p} = p(\mathbf{a}\,|\,\mathbf{x})$} is called GMNL if and only if \deleted{it does not admit any bilocal decomposition, i.e., } it cannot be expressed as a convex mixture of probability distributions local across \replaced{some nontrivial} bipartitions, i.e.,
%
\begin{equation}
    p(\mathbf{a}\,|\,\mathbf{x}) \neq \sum_k p_k \sum_{\lambda} r_{\lambda}^{(k)}\, p_{S_k\,|\,\bar{S}_k}(\mathbf{a}\,|\,\mathbf{x}, \lambda),
\end{equation}
%
where $p_k \geqslant 0, \ \sum_k p_k = 1$, and 
\deleted{where for each $k$, the distributions} $p_{S_k\,|\,\bar{S}_k}(\mathbf{a}\,|\,\mathbf{x}, \lambda)$ 
\added{denotes a probability distribution which is local with respect to the bipartition $S_k|\bar{S}_k$.}

admit a local hidden variable model \textbf {(Remik: This is not defined)} across the bipartition $S_k\,|\,\bar{S}_k$. Here, $\lambda$ denotes the hidden variable, and $r_{\lambda}^{(k)}$ is the probability distribution over these hidden variables for the bipartition indexed by $k$, satisfying $r_{\lambda}^{(k)} \geqslant 0$ and $\sum_{\lambda} r_{\lambda}^{(k)} = 1$. The variable $\lambda$ accounts for any classical correlations or shared randomness among parties within the bipartition.
\end{defn}
\fi

To detect that $\mathbf{p}$ is genuinely multipartite nonlocal one can use Bell inequalities. Recently, Curchod \textit{et al.}~\cite{curchod} introduced a general method for constructing families of Bell inequalities that witness GMNL for any number of parties. The approach is based on a two-party Bell inequality used as a seed to generate multipartite Bell inequalities. Violation of the resulting inequalities allows one to certify the presence of GMNL in $\mathbf{p}$. A particular example of such an inequality reads \cite{curchod}:
\begin{equation}
    \sum_{i=1}^{n-1}\sum_{j>i}^n I_{\mathbf{0}|\mathbf{0}}^{1,j}\leqslant (n-2)p(\mathbf{0}|\mathbf{0})\label{BI_detecting_GMNL},
\end{equation}
where $\mathbf{0} \equiv \underbrace{00\ldots0}_{n\,\text{times}}$, and the subscript $j$ refers to party $A_j$, and
\begin{align}\label{LiftedCHSH}
I_{\mathbf{0}|\mathbf{0}}^{1,j}=p(\mathbf{0}|\mathbf{0})&-p(10\ldots 0|10\ldots 0)\nonumber\\
&-p(0\ldots 01_j0\ldots 0|0\ldots 01_j0\ldots 0)\nonumber\\
&-p(\mathbf{0}|10\ldots 01_j0\ldots 0)
\end{align}
is the CHSH Bell expression \cite{CHSH1969} between parties $A_1$ and $A_j$ lifted to $n$ parties; the remaining $n-2$ parties' measurement and outcome is fixed as zero. (In the general case, one can choose an arbitrary party $A_i$ with $i\in [n]$, which may slightly alter the form of the inequality.) The seed inequality for this construction is the well-known CHSH inequality
\begin{equation}
    p(00|00)-p(01|01)-p(10|10)-p(00|11)\leqslant 0.
\end{equation}
It was shown in Ref.~\cite{curchod} that the inequality~\eqref{BI_detecting_GMNL} detects GMNL for an $n$-party scenario. Noticeably, these inequalities were recently improved in Ref.~\cite{Stachura_2024}. Namely the following tighter inequalities
are satisfied by any bilocal correlations:
\begin{equation}
    \sum_{j=2}^n I_{\mathbf{0}|\mathbf{0}}^{1,j}\leqslant (n-2)[p(\mathbf{0}|\mathbf{0})-p(1\mathbf{0}_{n-1}|1\mathbf{0}_{n-1})]\label{Improved},
\end{equation}
where $\mathbf{0}_{n-1}$ is a shorthand for a string of zeros of length $n-1$.

\section{\texorpdfstring{The $n$-qubit case}{}}\label{sec:main-result}

We are now in a position to present the main result which is a proof that any GME state belonging to a certain class of multipartite highly symmetric states is also GMNL.

A symmetric subspace $\mathrm{Sym}((\mathbb{C}^2)^{\otimes n})$ of the $n$-qubit Hilbert space~\cite{Marconi2025} is spanned by the so-called Dicke states
\begin{equation}
    |D^n_k\rangle = \frac{1}{\sqrt{\binom{n}{k}}} \sum_{\substack{x \in \{0,1\}^n \\ |x| = k}} |x\rangle,
\end{equation}
where $|x\rangle = |x_1 x_2 \dots x_n\rangle$ and $|x| = \sum_i x_i$~\cite{Dicke54}. Consider an arbitrary $n$-qubit pure state
\begin{equation}
    |\psi\rangle = \sum_{i_1,\dots,i_n} h_{i_1 \dots i_n} |\psi_{i_1}^{(1)}\rangle \dots |\psi_{i_n}^{(n)}\rangle, \label{generic-state}
\end{equation}
where each $|\psi_{i_k}^{(k)}\rangle \in \mathbb{C}^2$, and the coefficients $h_{i_1 \dots i_n} \in \mathbb{C}$ are complex amplitudes characterizing the state.  

Now, let us consider a generic $n$-qubit state that is symmetric under arbitrary permutations of a selected subset of $n-1$ parties. Without loss of generality, we may take this subset to consist of the last $n-1$ subsystems, with the remaining non-symmetric qubit being the first one. In other words, we consider $n$-qubit pure states from the space $\mathbb{C}^2 \otimes \mathrm{Sym}((\mathbb{C}^2)^{\otimes(n-1)})$, which can be expressed in the following way:
\begin{equation}
    |\psi\rangle = \sum_{k=0}^{n-1} \Big(h_k|0\rangle + h_k'|1\rangle\Big) \otimes |D^{n-1}_k\rangle,
    \label{main state}
\end{equation}
where $h_k$,$h_k' \in \mathbb{C}$ and $\sum_k (|h_k|^2 + |h_k'|^2) = 1$.

\begin{thm}
For any GME $n$-party pure state that is symmetric under permutations of $n-1$ parties (e.g., $A_2$ through $A_n$), there exist local measurements yielding correlations that violate inequality (\ref{Improved}), thus demonstrating the state to be GMNL.
\end{thm}

\begin{proof}
 We shall use the Bell inequality~\eqref{Improved} which can be simplified to
\begin{align}
    p(\mathbf{0}|\mathbf{0})&-\sum_{i=1}^np(0\ldots 01_i0\ldots 0|0\ldots 01_i0\ldots 0)\nonumber \\
    &- \sum_{i=2}^np(\mathbf{0}|10\ldots 01_i0\ldots 0) \leqslant 0.
\end{align}
Analogously to~Ref.~\cite{curchod}, we start by choosing parties $A_2$ to $A_n$ to perform the same projective measurements, meaning that $\langle m^{(2)}_{a_2|x_2}| =\ldots=\langle m^{(n)}_{a_n|x_n}|$, for every $a_2=\ldots=a_n$ and $x_2=\ldots=x_n$. This, together with the invariance of the state under permutation of any $2$ parties from $A_2$ to $A_n$, implies that the probability $p(a_1\ldots a_n|x_1\ldots x_n)$ also does not change under such permutations, and hence the above Bell expression simplifies to
\begin{align}\label{Catalonia}
     p(\mathbf{0}|\mathbf{0})&-p(10\ldots 0|10\ldots 0)\nonumber\\
     &- (n-1)p(010\ldots 0|01\ldots 0)\nonumber\\
     &- (n-1) p(\mathbf{0}|110\ldots 0) \leqslant 0.
\end{align}
We will now show that we can always find measurements to violate (\ref{Catalonia}) by satisfying conditions
\begin{align}
   p(10\ldots 0|10\ldots 0) &=0, \label{hardy condition 1} \\
   p(010\ldots 0|010\ldots 0)&=0, \label{hardy condition 2} \\
p(\mathbf{0}|110\ldots 0)&=0, \label{hardy condition 3}\\
p(\mathbf{0}|\mathbf{0})&>0,
\label{hardy condition 4}
\end{align}
which correspond to Hardy's paradox \cite{Hardy} between parties $A_1$ and $A_2$ lifted to an $n$-partite scenario.
Now, let us define the measurements for parties $A_2$ to $A_n$ conditioned on obtaining outcomes $a_2=\ldots=a_n=0$ for their input choice $x_2=\ldots=x_n=0$ as
\begin{equation}
    \langle m_{0|0}^{(j\geq2)}| \equiv \cos\alpha\,\langle 0| + \sin\alpha\,\langle 1|
    \label{m00symmetric}
\end{equation}
for some $\alpha\in[0,2\pi[$. Upon performing the measurements $\langle m_{0|0}^{(j)}|$ $(j=3,\ldots,n)$ on the state $|\psi\rangle$, the post-measurement state shared by $A_1$ and $A_2$ is of the form:
\begin{equation}    |\psi_2\rangle=b_1|00\rangle+b_2|01\rangle+b_3|10\rangle+b_4|11\rangle.
\end{equation}
Due to the fact that the $n-2$ measurements are of the form~(\ref{m00symmetric}) each component of the initial state contributes to a coefficient $b$ accordingly, so that it contains a mixture of sine and cosine factors, where the exponents reflect the number of $|1\rangle$ and $|0\rangle$ kets among the measured qubits, respectively. This contribution is multiplied by the appropriate parameter $h_k$ or $h_k'$, where $k$ equals the number of sine terms if the second unmeasured qubit is $|0\rangle$, and equals that number plus one if it is $|1\rangle$. The cosine exponent then complements the sine exponent so that their sum equals the total number of $n-2$ measured qubits.%, i.e., $n-2$.

Each such component of $b_j$ appears as many times as there are ways to distribute $k$ (or $k-1$) ones among the $n-2$ measured positions, leading to the following expressions:
\begin{align}
    \allowdisplaybreaks
    b_1&=\sum_{k=0}^{n-2} \binom{n-2}{k} h_k \cos^{n-k-2}\alpha \,\sin^{k}\alpha,\label{b1_sum}\\
    b_2&= \sum_{k=1}^{n-1} \binom{n-2}{k-1} h_k \cos^{n-k-1}\alpha\, \sin^{k-1}\alpha,\label{b2_sum}\\
    b_3&=\sum_{k=0}^{n-2} \binom{n-2}{k} h'_k \cos^{n-k-2}\alpha \,\sin^{k}\alpha,\label{b3_sum}\\
    b_4&=\sum_{k=1}^{n-1} \binom{n-2}{k-1} h'_k \cos^{n-k-1}\alpha\, \sin^{k-1}\alpha.\label{b4_sum}
\end{align}
The sums for $b_2$ and $b_4$ exclude terms with lower $k$, as the second position of the relevant basis states always contains at least one $|1\rangle$.

It is not difficult to observe that the state $|\psi_2\rangle$ is entangled if and only if $b_1b_4-b_2b_3 \neq 0$.
Knowing the form of $b_i$ coefficients, this condition can be expanded into
\begin{widetext}
\begin{align}
    b_1b_4-b_2b_3 &= \sum_{k=0}^{n-2} \sum_{l=1}^{n-1} (h_kh_l'- h_k'h_l) \binom{n-2}{k} \binom{n-2}{l-1} \cos^{2n-3-k-l}\alpha \sin^{k+l-1}\alpha\\
     & = \cos^{2n-4}\alpha\sum_{k=0}^{n-2} \sum_{l=0}^{n-2} (h_kh_{l+1}'- h_k'h_{l+1}) \binom{n-2}{k} \binom{n-2}{l}  \tan^{k+l}\alpha\label{factor_cosine2},
\end{align}
\end{widetext}
where Eq. \eqref{factor_cosine2} is a result of factoring out the part $\cos^{2n-4}\alpha$ and shifting the index. Clearly, these operations do not affect the values of $\alpha$ for which the expression \eqref{factor_cosine2} equals zero. In particular, it directly follows from Eq. (\ref{factor_cosine2}) that it vanishes for $\alpha=\pi/2$, $\alpha=3\pi/2$, and therefore in further considerations we need to exclude those values.

Now, we introduce $m=k+l$ and we can rewrite the expression as a polynomial in $\tan\alpha$ variable:
\begin{equation}
    b_1b_4-b_2b_3= \cos^{2n-4}\alpha\sum_{m=0}^{2n-4} C_m  \tan^{m}\alpha. 
    \label{concurrence polynomial}
\end{equation}
with
\begin{widetext}
\begin{equation}
    C_m=\sum_{k=\max\{0,m-n+2\}}^{\min\{n-2,m\}} (h_kh_{m-k+1}'- h_k'h_{m-k+1}) \binom{n-2}{k} \binom{n-2}{m-k}.\label{Cm-coefficients}
\end{equation}
\end{widetext}

As long as (\ref{concurrence polynomial}) is a nonzero polynomial, $|\psi_2\rangle$ remains non-maximally entangled over continuous ranges of $\alpha$. For (\ref{concurrence polynomial}) to be identically zero for continuous values of $\alpha$, all coefficients $C_m$ must vanish. In Appendix~\ref{app:C} we prove that
$
    \forall_m : C_m=0 \iff \forall_k : h_k=\lambda h_k',
$
for some $\lambda$, which implies that the original state~(\ref{main state}) would be biseparable with respect to the bipartition between the first qubit and the rest. This, however, contradicts the initial assumption that the state is genuinely multipartite entangled (GME). Then the polynomial~(\ref{concurrence polynomial}) can only be 0 for at most $2n-4$ discrete values of $\alpha$. Since $\alpha$ is continuous, we choose to avoid those values, as well as the previously excluded values $\alpha=\pi/2$, $\alpha=3\pi/2$, to obtain a non-maximally entangled state $|\psi_2\rangle$.

Following Ref.~\cite{Hardy}, for any pure, non-maximally entangled state 
$|\phi_\theta\rangle = \cos\theta|00\rangle + \sin\theta|11\rangle$ with $\theta\in]0,\pi/4[$,
we can find a one parameter family of measurements $\langle m^{(1)}_{a_1|x_1}|$, $\langle m^{(2)}_{a_2|x_2}|$ leading to Hardy's paradox. It was shown in~\cite{curchod} how such measurements can be constructed for the state $|\phi_\theta\rangle$. Since we have already demonstrated that $|\psi_2\rangle$ is non-maximally entangled, it can also be written as $|\phi_\theta\rangle$, implying that the measurements found in~\cite{curchod} would apply after an appropriate change of basis. This, in turn, leads to a violation of inequality~(\ref{Catalonia}), thus certifying genuine multipartite nonlocality (GMNL).
\end{proof}
We note that these measurements can be written directly, without any basis transformation. By setting~(\ref{m00symmetric}), the conditions~(\ref{hardy condition 1}–\ref{hardy condition 3}) yield (up to normalization) the following measurement vectors:
\begin{widetext}
\begin{equation}
        \langle m_{0|0}^{(1)}|\propto(b_3(b_1c_1 + b_3c_2)^*+b_4(b_2c_1 + b_4c_2)^*)\langle 0|-(b_1(b_1c_1 + b_3c_2)^*+b_2(b_2c_1 + b_4c_2)^*)\langle 1|
\end{equation}
\end{widetext}
\begin{align}    
    %\langle m_{0|0}^{(1)}|&\propto(b_3(b_1c_1 + b_3c_2)^*+b_4(b_2c_1 + b_4c_2)^*)\langle 0|\nonumber\\&-(b_1(b_1c_1 + b_3c_2)^*+b_2(b_2c_1 + b_4c_2)^*)\langle 1|,\\
    \langle m_{0|1}^{(1)}|&\propto c_1\langle 0|-c_2\langle 1|,\\
    \langle m_{0|1}^{(2)}|&\propto(b_2c_1 + b_4c_2)\langle 0|-(b_1c_1 + b_3c_2)\langle 1|,\\
    \langle m_{1|1}^{(1)}|&\propto c_2^*\langle 0|-c_1^*\langle 1|,\\
    \langle m_{1|1}^{(2)}|&\propto (b_1c_1+b_3c_2)^*\langle 0|-(b_2c_1+b_4c_2)^*\langle 1|,
\end{align}
where 
$c_1=b_1^*\cos\alpha + b_2^*\sin\alpha$ and 
$c_2=b_3^*\cos\alpha + b_4^*\sin\alpha$.

Finally, the condition~\eqref{hardy condition 4} must be satisfied since the state $|\psi_2\rangle$ is non-maximally entangled and thus it admits Hardy's construction.

\section{Conclusions and Future Prospects}\label{sec:conclusion}

In this work, we extend the analysis of genuine multipartite nonlocality in multiqubit systems. Building upon the results of~\cite{Stachura_2024}, we establish, by means of elementary algebraic arguments, that every genuinely multipartite entangled state of $n$ qubits, for which $n-1$ parties are permutation-invariant, necessarily exhibits genuine multipartite nonlocality. The certification of GMNL is guaranteed by the violation of a suitably chosen Bell inequality, which admits a simplified form due to the imposed symmetry of the system. One possible approach to the general case is to gradually relax the symmetry assumptions. Numerical simulations support the general conclusion, suggesting that the result holds true for arbitrary $n$-qubit configurations, however, a formal proof is still lacking and remains a subject for future investigation.

\vspace{-0.2in}
\section{Acknowledgments}
\vspace{-0.1in}
We acknowledge discussions with Owidiusz Makuta and Arturo Konderak. This work is supported by the National Science Centre (Poland) through the SONATA BIS project No. 019/34/E/ST2/00369, and  funding from the European Union's Horizon Europe research and innovation programme under grant agreement No 101080086 NeQST.

\appendix

\vspace{-0.1in}
\section{\texorpdfstring{Relation between the state parameters}{}}
\label{app:C}
If $C_m$ defined in~\eqref{Cm-coefficients} vanishes for all $m$ then it is sufficient and necessary condition for the parameters $h_k$ and $h_k'$ of the state~\eqref{main state} to be linearly dependent.

\begin{proof}
Let us now consider the first two $C_m$ coefficients. Assuming $n\geq3$ we have:
    $C_0=h_0h_1'-h_0'h_1$ and
    $C_1=(h_0h_2'-h_0'h_2)(n-2)+(h_1h_1'-h_1'h_1)(n-2)$.
From $C_0=0$, we obtain
    $h_0=\lambda h_0'$ and
    $h_1=\lambda h_1'$
for some constant $\lambda$. Substituting this into $C_1=0$ yields $h_2=\lambda h_2'$. Now, assume there exists an index $r$ such that
\begin{equation}
    h_i=\lambda h_i' \quad \text{for}\quad i\leq r\leq n-2.
    \label{linearity assumption}
\end{equation}
It is clear that $C_m=0$ for $m<r$. Consider
\begin{equation}
    C_r=\sum_{k=\max\{0,r-n+2\}}^{\min\{n-2,r\}} (h_kh_{r-k+1}'- h_k'h_{r-k+1}) \binom{n-2}{k} \binom{n-2}{r-k}.
\end{equation}
From (\ref{linearity assumption}) we infer that $(h_kh_{r-k+1}'- h_k'h_{r-k+1})=0$ for $k\neq0$, hence
\begin{equation}
    C_r = (h_0h_{r+1}'- h_0'h_{r+1}) \binom{n-2}{r}.
\end{equation}
By setting $C_r=0$ we also impose $h_{r+1}=\lambda h_{r+1}'$. Therefore, by induction,
    $\forall_m : C_m=0 \iff \forall_k : h_k=\lambda h_k'$, which confirms the claim.
\end{proof}

\bibliography{bibliography}

\end{document}